\documentclass[sn-mathphys, Numbered]{sn-jnl}

\usepackage{graphicx}%
\usepackage{aas_macros}%
\usepackage{multirow}%
\usepackage{amsmath,amssymb,amsfonts}%
\usepackage{amsthm}%
\usepackage{mathrsfs}%
\usepackage[title]{appendix}%
\usepackage{xcolor}%
\usepackage{textcomp}%
\usepackage{manyfoot}%
\usepackage{booktabs}%
\usepackage{algorithm}%
\usepackage{algorithmicx}%
\usepackage{algpseudocode}%
\usepackage{listings}%

\usepackage{longtable}

\graphicspath{{./}{figures}}

\newcommand{\nuc}[2]{$^{#1}$#2}

\begin{document}

\title[Astromers]{Astromers: Status and Prospects\footnote{This article is intended for unlimited release under LA-UR-24-20183.}}

\author*[1,3]{\fnm{G. Wendell} \sur{Misch}}\email{wendell@lanl.gov}

\author[2,3]{\fnm{Matthew R.} \sur{Mumpower}}\email{mumpower@lanl.gov}

\affil*[1]{\orgdiv{XTD-PRI}, \orgname{Los Alamos National Laboratory}, \orgaddress{\street{MS T086 TA-3 Bldg 2327}, \city{Los Alamos}, \postcode{87545}, \state{NM}, \country{USA}}}

\affil[2]{\orgdiv{Theoretical Division}, \orgname{Los Alamos National Laboratory}, \orgaddress{\street{MS B283 TA-3 Bldg 123}, \city{Los Alamos}, \postcode{87545}, \state{NM}, \country{USA}}}

\affil[3]{\orgdiv{Center for Theoretical Astrophysics}, \orgname{Los Alamos National Laboratory}}

\abstract{
The extreme temperatures and densities of many astrophysical environments tend to destabilize nuclear isomers by inducing transitions to higher energy states which may then cascade to ground.
However, not all environments destabilize all isomers.
Nuclear isomers which retain their metastable character in pertinent astrophysical environments are known as astrophysically metastable nuclear isomers, or ``astromers''. 
Astromers can influence nucleosynthesis, altering abundances or even creating new pathways that would otherwise be inaccessible. 
Astromers may also release energy faster or slower relative to their associated ground state, acting as heating accelerants or batteries, respectively. 
In stable isotopes, they may even simply remain populated after a cataclysmic event and emit observable x- or $\gamma$-rays.
The variety of behaviors of these nuclear species and the effects they can have merit careful consideration in nearly every possible astrophysical environment.
Here we provide a brief overview of astromers past and present, and we outline future work that will help to illuminate their role in the cosmos.
}

\keywords{astromers, nuclear isomers, nucleosynthesis, nuclear structure, nuclear reactions, nuclear astrophysics}



\maketitle

\section{Introduction}
\label{sec:introduction}

Nuclides in astrophysics are connected in networks by creation and destruction mechanisms (transmutations) including proton \cite{Schatz1999, Wanajo2006, Cyburt2016} and neutron capture \cite{Seeger1965, Mathews1983, Rauscher2000ncap, Ando2006, Cowan2021}, $\beta$ decay and its variants \cite{Takahashi1983, CaballeroFolch2014, Lund2023}, $\alpha$ capture \cite{Gorres1992, Rauscher2000acap, Siess2004, Elhatisari2015}, fission \cite{Mumpower2022, Holmbeck2023, Roederer2023fis}, and so on.
Astrophysical nucleosynthesis---the collective processes and environments in which the elements of the universe are made---depends sensitively on transmutation rates, which in turn depend on both the surrounding thermodynamic conditions and the properties of nuclei.

Where sufficient information (experimental data and/or theoretical estimates) about the excited states of an isotope of interest is available, analyses generally use a Boltzmann distribution of energy levels to incorporate the thermal average $\langle P \rangle$ of a key physical property $P$, e.g. the $\beta$-decay rate \cite{fuller1980stellar, fuller1982astellar, fuller1982b, fuller1985stellar, oda1994rate, langanke2000shell, langanke2001rate, misch2018neutrinospectra}; note that $P$ for an individual state may also depend on the environment.
\begin{align}
    \langle P \rangle &= \frac{1}{G(T)} \sum\limits_i P_i (2J_i+1) e^{-E_i / k_B T} \label{eq:thermal_average}
\end{align}
The sum in Eqn.~\ref{eq:thermal_average} runs over nuclear energy levels enumerated by index $i$, $J$ is the level's angular momentum (``spin''), $E$ is its energy, and $T$ is the temperature. 
$G(T)$ is the partition function.

\begin{align}
    G(T) &= \sum\limits_i (2J_i+1) e^{-E_i / k_B T} \label{eq:partition_function}
\end{align}

Where neither data nor theory provide adequate excited state information, the ground state properties are typically taken as a stand-in \cite{Mumpower2016r}.
In relatively low-temperature scenarios where excited state populations are suppressed, this is typically a useful approximation.

\begin{figure}
    \centering
    \includegraphics[width=0.6\columnwidth]{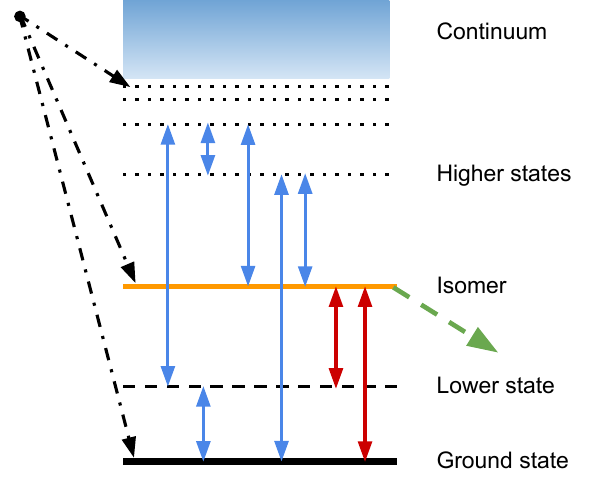}
    \caption{A schematic of the discrete energy levels in a nuclear system with an isomer (orange). The isomer has relatively slow (red) transitions to lower lying states. Relatively faster transitions (blue) can be enabled via thermal population. Solid arrows show electromagnetic transitions while a dashed-dotted (black) arrows indicate population from another nucleus. The isomer may be depopulated via a competing process (green dashed arrow). }
    \label{fig:isomer_levels}
\end{figure}

Nuclear isomers---excited states that live longer than the typical picoseconds or femtoseconds---call into question the validity of either approach \cite{Hahn1921, Dracoulis2016, Walker2020review}.
By way of illustration, Figure \ref{fig:isomer_levels} shows some essential characteristics and notional transmutations of a nucleus with an isomer.
In this nucleus, the isomer is the second excited state (orange), and there is an assortment of higher-lying states before the continuum.
The first excited state is not an isomer.

The vertical arrows represent internal transitions between energy levels; they point both up and down to indicate that the surroundings can induce upward transitions in addition to the spontaneous and stimulated downward transitions.
Blue arrows delineate typical fast transitions.
Isomers exist because of quantum mechanical mismatches between themselves and lower-lying states that hinder transitions \cite{Walker1999energy}; the hindered transitions are shown in red.

Finally, nuclear transmutations create (black dash-dot) and destroy (green dash) the nuclear species; note that the ground state may also be destroyed, but this is omitted from Figure \ref{fig:isomer_levels} for clarity.
This brings us to the question of how isomers might influence nucleosynthesis.
On the one hand, the hindered transitions slow thermal equilibration, and transmutations could destroy the isomer faster than it can be replenished.
On the other hand, sufficiently high temperatures will populate higher states that in turn rapidly populate the isomer, ensuring thermal equilibrium.
Further complicating the situation, transmutations that create the isotope might cascade into or directly feed the isomer.
The hindered transitions in this case would invalidate the use of ground state properties.
Unless, of course, it were hot enough to rapidly drive the isomer to ground via higher states.
This is the essence of the conundrum of isomers in astrophysical nucleosynthesis.


\section{History}

This timeline is by no means complete, but it highlights several milestones in the study of isomers in astrophysics.

The problem of nuclear isomers in astrophysical environments was first identified and tackled by \citet{Solomon1978stellar} in 1978.
They were examining a bottleneck at $A=45$ in silicon burning as nucleosynthesis flows from \nuc{28}{Si} to the iron group nuclei. 
The key reaction \nuc{45}{Sc}($p,\gamma$)\nuc{46}{Ti} presented a particular challenge.
Not only will \nuc{45}{Sc}'s lowest excited state (only 12.4 keV) have a high equilibrium population within the relevant temperature range of 100-500 keV, but also that state is an isomer with a half-life of 326 ms.
Thus, the ($p,\gamma$) cross section of that state must be included, and it is not obvious if its in-situ population would be the thermal equilibrium value or if proton captures would overwhelm internal transitions.
With a computed mean lifetime against proton capture of 1 ns, the question boiled down to how rapidly the ground state and isomer populations---produced by \nuc{42}{Ca}$(\alpha, p)$\nuc{45}{Sc} and \nuc{45}{Ti}$(n, p)$\nuc{45}{Sc}---converged to thermal equilibrium.
By following the populations of all known \nuc{45}{Sc} states as they transitioned amongst themselves, they found that the ensemble deviated from a Boltzmann distribution by less than 10\% in no more than 140 ps, concluding that thermal equilibrium was an adequate assumption in their application.

Building on this work, \citet{ward1980thermalization} developed an analytical formulation for the ground state/isomer thermal equilibration timescale as a function of temperature in nuclei with 1-2 non-isomeric excited states, then extended it to more realistic nuclei with many levels.
Focusing on \nuc{26}{Al}, they compared the equilibration time to nuclear transmutation rates in an assortment of astrophysical conditions where that isotope can play a role.
They found that in some situations, \nuc{26}{Al} would reach thermal equilibrium, while in other environments, the destruction timescale was much shorter than the thermalization timescale.

Using more complete data and large-scale nuclear shell model calculations to compute essential unmeasured quantities (transition matrix elements and $\beta$-decay rates), \citet{coc1999lifetimes} tackled \nuc{26}{Al} and \nuc{34}{Cl}.
In their approach, they performed a network calculation where the ``isotopes'' were nuclear energy levels.
From this, they computed effective lifetimes of the ground state and isomer for each nuclide, which in practice is the quantity required for nucleosynthesis simulations.

\citet{Gupta2001} created a general treatment of isomers in astrophysics by constructing particular ensembles of excited states.
They demonstrated that non-isomeric excited states rapidly achieve a Boltzmann-like equilibrium with the ground state and isomer \emph{independently}.
This led to an interpretation of excited states as being associated with ground or the isomer proportionally to the probability that it transitions to one or the other.
The ensembles then consist of a single long-lived state (ground state or isomer) and a modified Boltzmann distribution where each excited state's contribution in Eqn. \ref{eq:thermal_average} is further weighted by its association with that long-lived state.
With this construction, the ground state and isomer are treatable as truly distinct species that undergo their own transmutations and transitions between one another.
\citet{Gupta2001} also conceptualized the nuclear levels and transitions as a directed graph and used Dijkstra's algorithm for finding shortest paths \cite{dijkstra2022note} to identify key internal transitions.

Recently, \citet{Misch2021astromers} found a precise formulation for computing thermally mediated ground state $\leftrightarrow$ isomer transitions via intermediate excited states that minimizes physical assumptions.
In that approach, transition probabilities between all nuclear levels enable the calculation of effective transition rates between the long-lived states with a single matrix inversion; no assumptions about excited state populations are required.
By defining an effective transition as any chain of transitions that starts in one long-lived state and ends in another, they were able to generalize to nuclides with multiple isomers.
As with \citet{Gupta2001}, the rates can then be tabulated as functions of temperature, and each of possibly multiple isomers may be treated as separate species from the ground state in nucleosynthesis calculations.

The preceding history focused on the development of techniques, initiated by Solomon and Sargood's \cite{Solomon1978stellar} needs for nucleosynthesis studies around $A\approx 45$.
But naturally, that development would not have continued if isomers didn't seem to be important elsewhere, as implied by the fact that many of the cited works spotlighted \nuc{26}{Al}.
Indeed, many efforts were driven in large part by \nuc{26}{Al}'s importance in meteoric isotopic abundances \cite{lee1977aluminum} and $\gamma$-ray astronomy \cite{Diehl1995}.
Over the years, other isomers have drawn scrutiny from the nuclear astrophysics community.

Beyond the techniques for quantifying the behavior of isomeric species, other research into isomers in astrophysics includes theoretical \cite{Runkle2001, Banerjee2018, Richter2020shell} and experimental \cite{Reed2012, Kahl2017isobeam, Chipps2018, Fan2023mo} nuclear inputs.
Software tools for computing in-situ rates have been constructed \cite{Reifarth2018, Misch2021astromers, Tannous2023}, and isomers have been included in nucleosynthesis codes \cite{Iliadis2011effects, Sprouse2022}. 
The breadth of nucleosynthesis environments under examination for isomer effects has grown, and with it, the isomers under investigation \cite{Takahashi1983, Meyer2001iso, Fujimoto2020, Richter2020shell, Misch2021astror, Misch2021sensitivity}. 
Now, highly developed methods, modern nucleosynthesis codes, and ever-improving experimental capabilities have converged with expanding recognition of the importance of isomers over the last several years to inspire a burst of interest in this burgeoning field \cite{Aprahamian2005long, Norman2023}.


\section{What makes an isomer an astromer?}
\label{sec:astromer}

Simply put, an astromer is a nuclear isomer that retains its metastable character in an astrophysical environment. 
High temperatures and high densities tend to drive internal transitions both up and down between nuclear levels, pushing the ensemble of states to an equilibrium Boltzmann distribution. 
However, an isomer, by definition, is at least partially isolated by inherently slow transitions. 
Sufficiently strong thermal effects can overcome this isolation, and the interplay of metastability and the surrounding environment sets the isomer's behavior.

``Metastable character'' in this context means that the excited state resists internal transitions to such an extent that the ensemble is far from a Boltzmann distribution for sufficiently long that the nuclear species exhibits non-thermal-equilibrium behavior. 
If an isomer undergoes nuclear transmutation more rapidly than internal transitions, thermal equilibrium may not be attainable. 
Or perhaps a thermal population of an excited state ``freezes out'' as an astrophysical event cools (the transition rates become so slow that a non-thermal distribution remains as the temperature drops).
In these situations, the nucleus may not be treatable only with ground state properties, nor may properties derived from a thermal Boltzmann distribution be safely assumed.
The ground state and isomer must be considered as separate species; this makes an isomer an astromer.

\begin{figure*}
    \centering
    \includegraphics[width=80 mm]{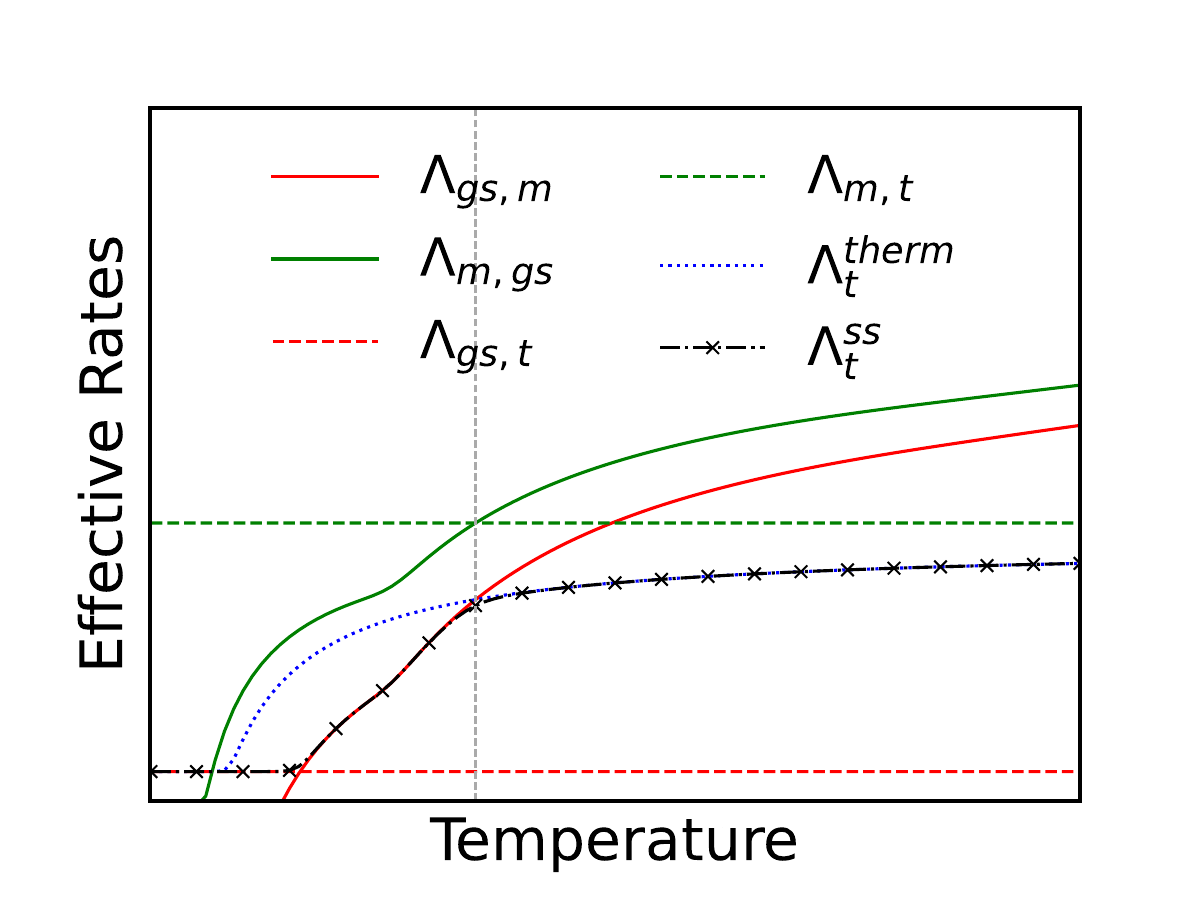}
    \caption{A schematic astromer diagram where the isomeric state $m$ transmutes (green dashed) faster than the ground state $gs$ (red dashed).
    The solid lines are effective ground $\leftrightarrow$ isomer transition rates.
    The fact that this isomer can be an astromer is evidenced by comparing the thermal equilibrium (blue dotted) and steady state (black dash-dot with crosses) transmutation rates:
    below the thermalization temperature associated with the transmutation $t$ (vertical line), the isotope clearly does not transmute as though it were in thermal equilibrium, and the isomer is an astromer.
    Additional reaction channels or decays further complicate this simple picture by introducing more contributions to the transmutation rates. See text for details.}
    \label{fig:astromers_schematic}
\end{figure*}

Figure \ref{fig:astromers_schematic} shows a schematic ``astromer diagram'' that illustrates astromer behavior.
The solid lines indicate ground state $\leftrightarrow$ isomer transitions, and broken lines show transmutation rates.
Red lines have ground ($gs$) as the initial state, and green lines have the isomer ($m$) as the parent.
The blue line is the nuclide's transmutation rate if the states are in thermal equilibrium (if the isomer acts like a normal excited state), and the black line shows the transmutation rate when ground and the isomer are in a \emph{steady state} equilibrium in which the population ratios are unchanging. 
In this example, the isomer is less stable (has a faster transmutation rate) than ground.

The key feature to observe is that there is a temperature regime where the thermal equilibrium and steady state rates do not overlap.
Thermalization, because it is driven by internal transitions, occurs when the transition rate out of each state dominates its transmutation rate, that is, when the process that effects thermal equilibrium runs faster than processes that would disrupt it.
In Fig. \ref{fig:astromers_schematic}, for example, observe that where the isomer transition rate (green solid) exceeds the isomer transmutation rate (green dashed), the thermal equilibrium and steady state transmutation rates converge. 
This defines the thermalization temperature (highlighted by the vertical line), above which thermal equilibrium may be safely assumed. 
But below, transmutations disrupt the Boltzmann distribution faster than internal transitions restore it, and thermal equilibrium does not hold. 
The isomer is then an astromer, and it must be treated separately from ground. 

Note that Figure \ref{fig:astromers_schematic} has been simplified to be pedagogical.
First of all, it shows only a single isomer, though an isotope can have two or more isomers.
Second, in astrophysical environments, there are often many transmutation processes that are active at different temperatures, densities, and material compositions.
The sum of their rates in a particular situation defines an isotope's thermalization temperature, not necessarily a single rate, and each transmutation channel is usually quite sensitive to the environment.
Therefore, assessing an isomer's astromeric nature demands careful consideration of environmental effects on transmutation rates, as there can exist a hierarchy of temperatures where a nucleus and its long-lived states may fail to thermalize.

\begin{figure*}[ht]
    \includegraphics[width=\textwidth]{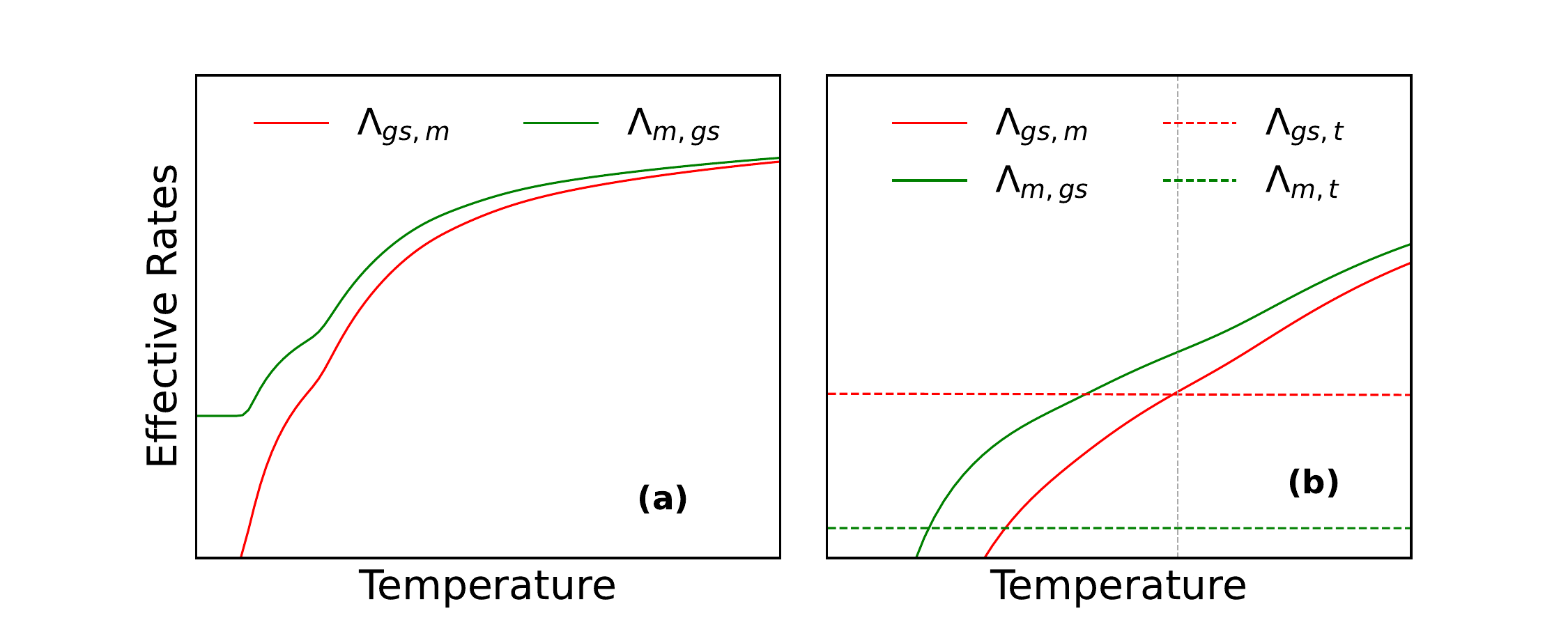}
    \caption{Schematic astromer diagrams of a) a stable isotope with an isomer and b) an isotope in which the isomer is more stable against transmutations than the ground state.
    In both cases, when the surroundings are sufficiently cool, an outsize proportion of the nuclei can persist in the isomer, which is then an astromer.}
    \label{fig:other_astromers}
\end{figure*}

Isomers can occur in stable nuclei, and it is even possible for isomers to be more stable than their corresponding ground states.
Fig. \ref{fig:other_astromers} shows examples in the left and right panels, respectively.
The latter case is similar to Fig. \ref{fig:astromers_schematic}, but with the ground and isomer transmutation rates reversed.
When this happens, it is possible to have more than the expected population in the isomer, because the ground state is destroyed faster.
Isomers in stable nuclei may persist after explosive nucleosynthesis has concluded.
Although there is no well defined thermalization temperature, the transition rates slow as the environment cools, and the isomer populations will ``freeze'' in place out of thermal equilibrium.
As these astromers relax to the ground state, the nucleus will emit x- or $\gamma$-rays.
With this variety of behaviors, in some astrophysical environments astromers will change nucleosynthetic pathways, in others they will affect energy release, and in still others they may give an observable electromagnetic signal.

The final component of understanding what makes an isomer an astromer is a detailed study of the role of non-isomeric excited states.
Nuclear states may be modeled as a directed graph in which the vertices (nodes) are the states, and the edges are transitions \cite{Gupta2001, Misch2021astromers, Ghosh2021dgraph}.
When the edges are weighted by the transition rates, pathfinding algorithms can rank which chains of intermediate states and associated transitions (paths) most contribute to the effective ground $\leftrightarrow$ isomer transition rates.

\begin{figure*}[ht]
    \centering 
    \includegraphics[width=90mm]{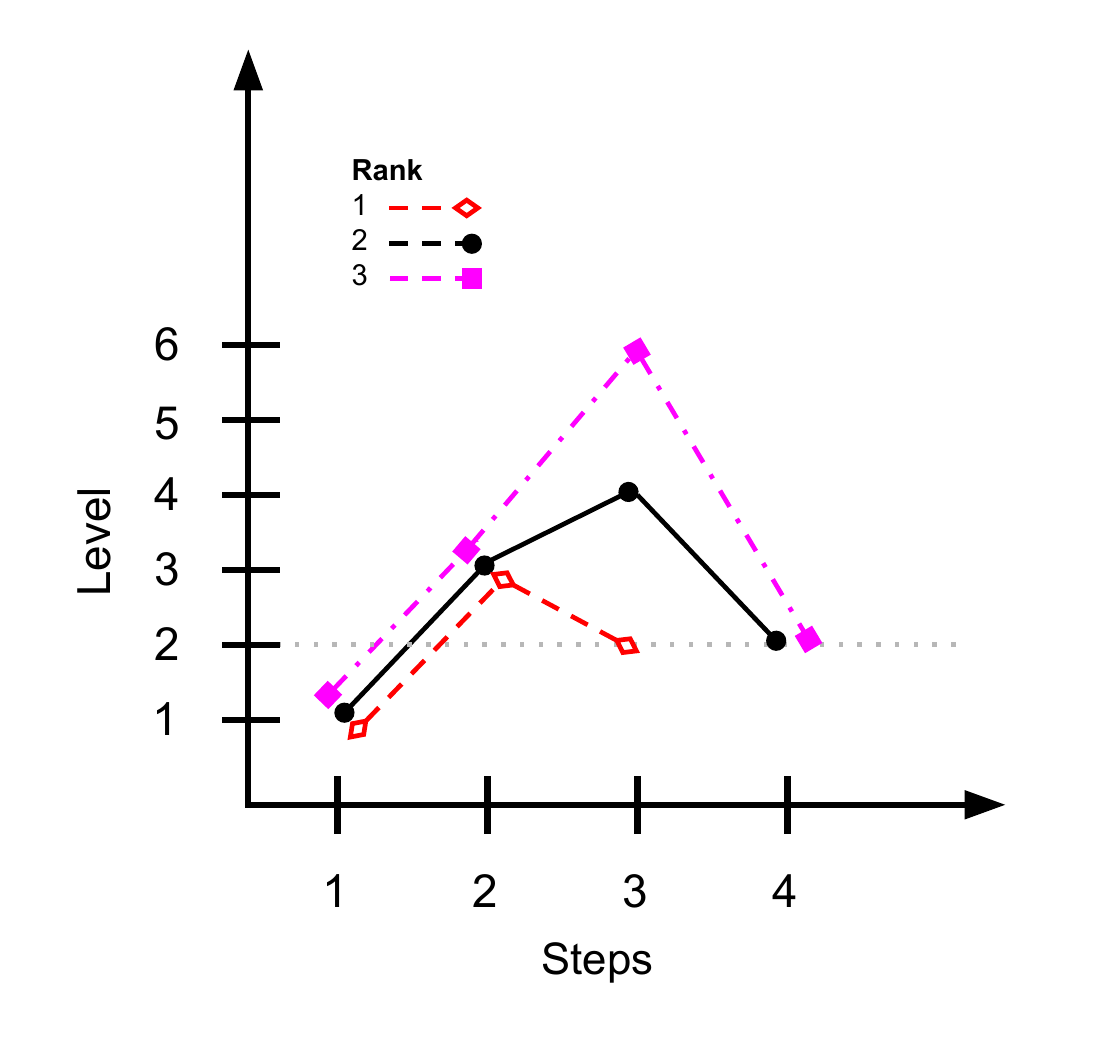}
    \caption{The most efficient paths (red, then black, then magenta) between the ground state (1) and isomeric state (2) for fixed environmental conditions. The connection between excited states is an essential ingredient in determining the astromeric nature of a nuclear level. See text for details. }
    \label{fig:isomer_path}
\end{figure*}

Figure \ref{fig:isomer_path} highlights the top three paths in a hypothetical nucleus; in general, which paths are most efficient is a function of the environment. 
On the vertical axis, the level numbers are in increasing order of state energy. 
The number of steps along the path from the ground state (Level 1) to the isomer (Level 2) is shown on the horizontal axis. 
The path with the overall fastest transitions between the ground state and the isomer is three steps in length (dashed red).
Paths ranked two (solid black) and three (dash-dotted pink) take four steps each.
Fig. \ref{fig:isomer_path} shows paths from ground to the isomer, but the most efficient paths from the isomer to ground are the same traveled in reverse \cite{Misch2021astromers}.

Path efficiency is not set by the number of steps, however, and a longer path can often be more efficient than a shorter one. 
Notice the inflection points in the transition rate curves in Figs. \ref{fig:astromers_schematic} and \ref{fig:other_astromers}. 
Those are hallmarks of a new path becoming thermally accessible. 
In the case of Fig. \ref{fig:astromers_schematic} (on which Fig. \ref{fig:isomer_path} is based), the $1 \rightarrow 3 \rightarrow 2$ path dominates at lower temperatures but is overtaken by the $1 \rightarrow 3 \rightarrow 4 \rightarrow 2$ path at higher temperatures.
This is because the $4 \rightarrow 2$ transition naturally outpaces the $3 \rightarrow 2$ transition; once 4 can be adequately fed by thermal transitions, it contributes to a faster path.

Pathfinding makes clear which nuclear data must be improved in order accurately compute the rates, whether an excited state's energy or a missing transition rate that had to be estimated.
This technique facilitates targeted experiments and theoretical investigation for accurate modeling of influential astromers. 


\section{Isomers in the cosmos}
\label{sec:isocosmos}

A variety of thermodynamic conditions abound in the cosmos from slow nuclear burning taking place in the interior of stars to explosive environments like compact object mergers where the astrophysical rapid neutron capture ($r$ process) is thought to take place. 
Each of these events has temperature and density evolution that characterize the environment, bringing about various behaviors of nuclear isomers.
Isomer studies have frequently been performed for stable or near-stable nuclei because this is where the most nuclear data is available. 
More recently, studies have pushed towards understanding the behavior of nuclei farther from stability. 
In this section, we summarize the present work regarding research into isomers in astrophysical environments.

The most well known astromer is the 228 keV state in \nuc{26}{Al}, which is synthesized inside relatively massive stars.
\nuc{26}{Al} was the first radioisotope to be observed in cosmos \citep{Mahoney1982, Diehl1995, Lugaro200826al}.
The ground state decays with a half-life of $\sim$700 kyr and emits a characteristic 1809 keV $\gamma$-ray, so this nucleus can serve as a tracer of star formation in the galaxy \citep{norgaard198026, clayton198426, singh2012evaluation}.
In contrast, the isomer half-life is only $\sim$6 s \cite{Kondev2021}.
Considering only $\beta$-decay, the thermalization temperature is around 35 keV.
Below this temperature, the isomer decays faster than thermally moderated internal transitions can achieve a Boltzmann distribution, causing this state to become an astromer.
The importance of \nuc{26}{Al} is widely known, and research continues on its production and destruction \cite{lederer2021destructiona, lederer2021destructionb, temanson2023measurement, battino2023impact}.

The slow neutron capture process ($s$ process) produces roughly half the heavy elements above iron on the periodic table; see Refs.~\cite{Kappeler2011, Lugaro2023} for recent reviews. 
The weak component (up to mass number, $A\sim90$) is thought to occur in the helium shell of massive stars, and the main $s$ process is thought to occur in asymptotic giant branch (AGB) stars \cite{Meyer1994}. 
$S$-process studies have traditionally assumed a thermal population of isomers, which even then alter reaction rates \cite{Kappeler1993}.
In certain cases, more detailed treatments are warranted, with nonthermalized effects considered by \citet{Kappeler1989}. 

An interesting $s$-process case arises with the nucleus \nuc{85}{Kr}. 
Because of its short half-life of 4.5 hours (as compared to the months to years timescales of $s$-process nucleosynthesis), the 304.8 keV state in \nuc{85}{Kr} acts as important astromer and branch point \cite{Walter1986, Abia2001}. 
When \nuc{85}{Kr} is populated below its $\beta$-decay thermalization temperature, the isomer has an 80\% $\beta$-decay branch, creating \nuc{85}{Rb}. 
In contrast, the ground state has a half-life of nearly 11 years, giving ample opportunity to capture a neutron to become stable \nuc{86}{Kr}, which can then capture another neutron and undergo subsequent $\beta$ decay to \nuc{87}{Rb}.
Consequently, the branch point at \nuc{85}{Kr} in the $s$ process greatly impacts the \nuc{87}{Rb} cosmochronometer \cite{Beer1984} (a cosmochronometer is a nucleus which serves to determine the age of an astrophysical event). 
Interestingly, the main $s$ process environment oscillates between temperatures below and above the thermalization temperature, so the nucleus oscillates between astromeric and thermal behavior \cite{busso1999nucleosynthesis}. 
Recent measurements have improved the \nuc{87}{Rb} decay constant, indicating previous age estimates of stars or samples in geochemistry to be 2\% too young \cite{Nebel2011}. 
This development merits further study of \nuc{85}{Kr} nucleosynthesis, particularly with respect to the astromer. 

Other isomers also influence the $s$ process. 
The observed presence of \nuc{99}{Tc} on the surface of red giant stars suggests the isomer plays a significant role in reducing the thermal $\beta$-decay rate of the nucleus \cite{Merrill1952, Richards1982technetium}. 
Shell model calculation for the $\beta$-decay of this nucleus later confirmed this behavior \cite{Takahashi1986}. 
The branch point \nuc{148}{Pm} also contains an isomer that affects its $\beta$-decay rate \cite{Winters1986, Lesko1989}. 
The importance of tin and indium isotopes was studied in the late 1980s \cite{Beer1989tin}. 
The nucleus \nuc{113}{Cd} and its isomer at 263.5 keV is another branch point in the $s$-process and is influential to the production of Sn isotopes \cite{Hayakawa2021}. 
The short-lived K-isomer \cite{Ghorui2018highk} at 123 keV in \nuc{176}{Lu} serves as a thermometer for $s$-process nucleosynthesis \cite{Heil2008}; precise astrophysical lifetime measurements are critical for accurate modeling of this thermometer and its function as a cosmochronometer \cite{Doll1999, Soderlund2004, Hayakawa2023}. 

The $s$ process is believed to end with a cycle in which \nuc{210}{Bi} $\beta$ decays to \nuc{210}{Po}, which in turn $\alpha$ decays to \nuc{206}{Pb}. 
An alternative ending occurs if the 3-million-year \nuc{210}{Bi} isomer is populated and is an astromer. 
Instead of the ground-state $\beta$ decay to \nuc{210}{Po}, the isomeric state would capture a neutron, then $\alpha$ decay to \nuc{207}{Tl}. 
However, studies of the impact of this isomer suggest thermal effects wash out any influence it may have on the end of the $s$-process, so treatment as a single species seems to be sufficient \citep{Ratzel2004}.

As the timescale for nucleosynthesis processes become faster, there's the potential to populate more nuclei and more isomers farther from stability. 
Data is less complete for short-lived nuclei, and thus there are more uncertainties associated with these isomer studies. 

The rapid proton capture process ($rp$ process) occurs in explosive hydrogen burning in type I x-ray bursts \cite{Wallace1981}. 
There are many isomers in the neutron-deficient $rp$ process path \cite{Kankainen2005, Sun2005, Garnsworthy2009}.
The temperature can range between 0.1 and 10 GK ($\sim$ 8-1000 keV), so isomers will certainly be populated \cite{VanWormer1994, Fisker2008} and should be evaluated for their astromer properties.
The \nuc{92}{Nb} cosmochronometer has been found not to be impacted by its isomer \cite{Mohr2016}, and \nuc{96}{Ru} is unlikely to be produced in $rp$ process \cite{Lorusso2011}.
However, other isomers such as \nuc{24}{Al} and \nuc{38}{K} can create branch points, and studies of them are ongoing.

Numerous experimental efforts have provided essential data for these studies.
\nuc{24}{Al} isomer production has been constrained \cite{Gerken2021exp}, although further information on an excited state in \nuc{25}{Si} is needed to evaluate the \nuc{24}{Al}(p,$\gamma$)\nuc{25}{Si} reaction rate. 
The \nuc{38}{K} isomer influences the bottleneck \nuc{38}{K}(p,$\gamma$)\nuc{39}{Ca} reaction, and techniques have been developed for reliably producing and characterizing it \cite{Chipps2018}.
\nuc{96}{Ag} has been found to contain multiple isomers that could be astromers \cite{Becerril2011}, although their energies remain unknown.
Penning trap mass spectrometry and other methods have been used to measure isomer energies \cite{Kankainen2010} and to distinguish between ground and excited states relevant for $rp$-process nuclei \cite{Fallis2011, Kankainen2012, Yan2013}. 
Isomer data---including half-lives, $\gamma$-ray energy, and even new isomers---has been collected in the \nuc{100}{Sn} region \cite{Kaneko2008, Park2017, Hornung2020}. 

Rapid neutron capture process ($r$ process) nucleosynthesis also involves explosive astrophysical conditions in which the temperature and density evolve swiftly \cite{Horowitz2019}. 
Although isomers have been identified as being important in the $r$ process \cite{Kajino2019}, they have only recently been included in simulations.

\citet{Fujimoto2020} were the first to study $r$-process isomers. 
In their work they assumed direct population from the $\beta$-decay parents of select tin, antimony, and tellurium isomers, which might have a substantial impact on kilonova light curves due to the large difference in $\beta$-decay half-lives between the isomeric and ground states. 
These elements are among the most populated in an $r$-process event, residing in the second abundance peak (mass number $A=130$), and hence have an outsize impact on observables. 

The first work to show the dynamic (not pre-assumed) population of nuclear isomers in the $r$ process was Ref.~\cite{Misch2021astror}. 
\citet{Misch2021astror} showed that not only are nuclear isomers populated during the decay back to stability (after neutron capture has mostly run its course), but also a substantial number of those isomers are in fact astromers. 
Different isomer behaviors were identified, including some that act as batteries (which store energy, releasing it at later times relative to the ground state population), some accelerants (which speed up energy release relative to the ground state), and some which do not strongly affect energy release, but might give an observable signal in the $r$-process remnant. 

This study used data from the latest ENDF and ENSDF libraries to explore known isomers. 
Influential nuclei with missing data in these evaluations were isolated. 
For example, the first excited state in $^{128}$Sb was proposed to be an important accelerant astromer, but its energy was not well measured and its level structure information was sparse. 
In this case, the missing data prevented proper determination of the thermalization temperature. 
\citet{Hoff2023prl} recently measured the first excited state of \nuc{128}{Sb} and found it to be 43.9 keV, which is significantly higher than a previous estimate. 
This measurement and the companion theoretical work enabled evaluation of the thermalization temperature, estimated to be between 1 and 9 keV.
This is much higher than the temperature of the $r$ process during its production, and this nucleus is therefore always an astromer in the $r$ process. 
This work underlines the synergistic combination of theoretical and experimental effort required for the study of astromers. 
A natural next step for both theoretical and experimental astromer efforts for the $r$ process will be to focus on nuclei in the $N=126$ region due to the large abundances of those nuclei coupled with expected rampant isomerism. 

Using pathfinding between known energy levels in neutron-rich nuclei (see the discussion surrounding Fig.~\ref{fig:isomer_path}), \citet{Misch2021sensitivity} estimated the sensitivity of isomer behavior to individual nuclear transitions. 
For a wide swath of neutron-rich isomers that could be produced in the $r$ process, they computed ranges of transition rates and thermalization temperatures.
The ranges came from varying the rates of unmeasured internal transitions.
Their pathfinding algorithm identified which unmeasured transitions most affected thermalization temperatures.

One of the profound remaining mysteries regarding isomeric state production in the cosmos is the case of \nuc{180}{Ta} \cite{Shizuma2002}. 
The metastable state of this isotope (77.2 keV) has a half-life of greater than $10^{15}$ yr, while the ground state decays in 8 hr. 
Therefore, the isomer prevents \nuc{180}{Ta} from decaying away after its production, affording it the opportunity to survive to be the rarest primordial isotope on Earth. 
The production site of \nuc{180}{Ta} remains unclear \cite{Belic1999, Belic2002, Mohr2007, Hayakawa2010, Bisterzo2015}. 


\section{Future work}
\label{sec:future}

Computational developments in combination with the prevalence of modern experimental facilities capable of producing a range of atomic nuclei make the time ripe for studying nuclear isomers in astrophysics. 
Here we list at a high level ideas for future research directions. 

The prediction of isomeric states in short-lived nuclei is essential. 
Predictions serve as a road map for experimental studies and supplement nucleosynthesis calculations where data is lacking. 
The nuclear shell model's ability to make detailed calculations of excited state properties (e.g. energy, spin, transition/decay matrix elements) makes it an ideal tool for computing the relevant nuclear structure, including not only predicting the existence of isomers, but also in calculating transition and transmutation rates.
Spherical shell model codes such as NuShellX \cite{brown2014shell} and BIGSTICK \cite{johnson2013factorization, johnson2018bigstick} are effective for nuclei that are not heavily deformed.
In regions of the chart of nuclides with many deformed nuclei, a projected shell model works well \cite{Hara1995, Sun2006, Sun2008projected, Wang2016}.
Because shell models compute both level structure and rates, they provide a self-consistent picture of excited states in nucleosynthetic environments. 

Isomer measurements have expanded in recent years \cite{Zhang2019isomer, Masuda2019x, Sikorsky2020measurement, Liu2020isomeric, Walker2020properties, Manea2020iso, Orford2020isomer, Nesterenko2020isomer}, including isolating rapidly decaying states, spectroscopy analysis, and mass measurements.
Isomers have even been accessed via laser excitation \cite{Lakosi1993, Ma2022, Ong2023}.
More and more short-lived nuclei are becoming accessible at radioactive beam facilities \cite{Moon2014}, but even short-lived nuclei have been found to have relevant isomeric states \cite{Jungclaus2007}. 
Pure isomeric beams offer a coherent picture of excited states allowing for in-beam $\gamma$ spectroscopy \cite{Pfutzner1997, Rykaczewski1998, SantiagoGonzalez2018, Asher2018}.
Distinguishing ground and isomeric states is also indispensable; new mass measurements using MR-TOF allows for precision measurements that discriminate between ground state and isomeric state \cite{AyetSanAndres2019}. 

The direct population and transmutation of isomers in nuclear reactions is also a fruitful area of research. 
While relatively unexplored, neutron capture creating isomers can substantially impact $s$-process nucleosynthesis \cite{Nemeth1994}.
Isomer effects on proton capture in the $rp$ process and neutron capture in the $r$ process both remain uncharted.
In \nuc{26}{Al} and others, determining the cross sections on excited states helps to reduce nucleosynthesis uncertainties \cite{Lotay2022}. 
Additionally, fission may populate different excited states in the daughter fragments than $\beta$ decay does; the impact on $r$-process nucleosynthesis has not been investigated. 

Theoretical predictions and experimental measurements of ever more exotic nuclei enable their evaluation. 
The release of evaluations in consistently updated databases is essential for accelerating the pace of scientific discovery in astrophysics. 
Databases that come equipped with code-based interfaces or easy-to-parse file formats reduce the need for one-off scripts in research. 
The release of the atlas for nuclear isomers represents a step in this direction \cite{Jain2015atlas, Garg2023atlas}. 

It is instructive to analyze the current status of data (and lack their of) affecting isomers. 
Uncertain nuclear data which affects \textit{known} isomers is shown in Figure \ref{fig:isomer_measurements} and tabulated in the appendix (Table \ref{table:missing_data}). 
Isomers with unknown energies are highlighted in red. 
Penning trap mass measurements can differentiate these states from the ground state. 
Green indicates neutron-rich nuclides with missing transition rates between energy levels isolated from \citet{Misch2021sensitivity}; that study covers a small subset of all possibly important unknown transition rates in atomic nuclei. 
Missing and incomplete data on branching ratios associated with the non-electromagnetic decay of nuclear levels are shown in blue; this data is important to determine the feeding of the isomer versus the ground state. 
For many nuclei, multiple properties will be required, as indicated by purple shading; \nuc{195}{Pt} for the understanding of the $r$-process is one such case \cite{Misch2021astror}. 

\begin{figure*}[ht]
    \centering
    \includegraphics[width=\textwidth]{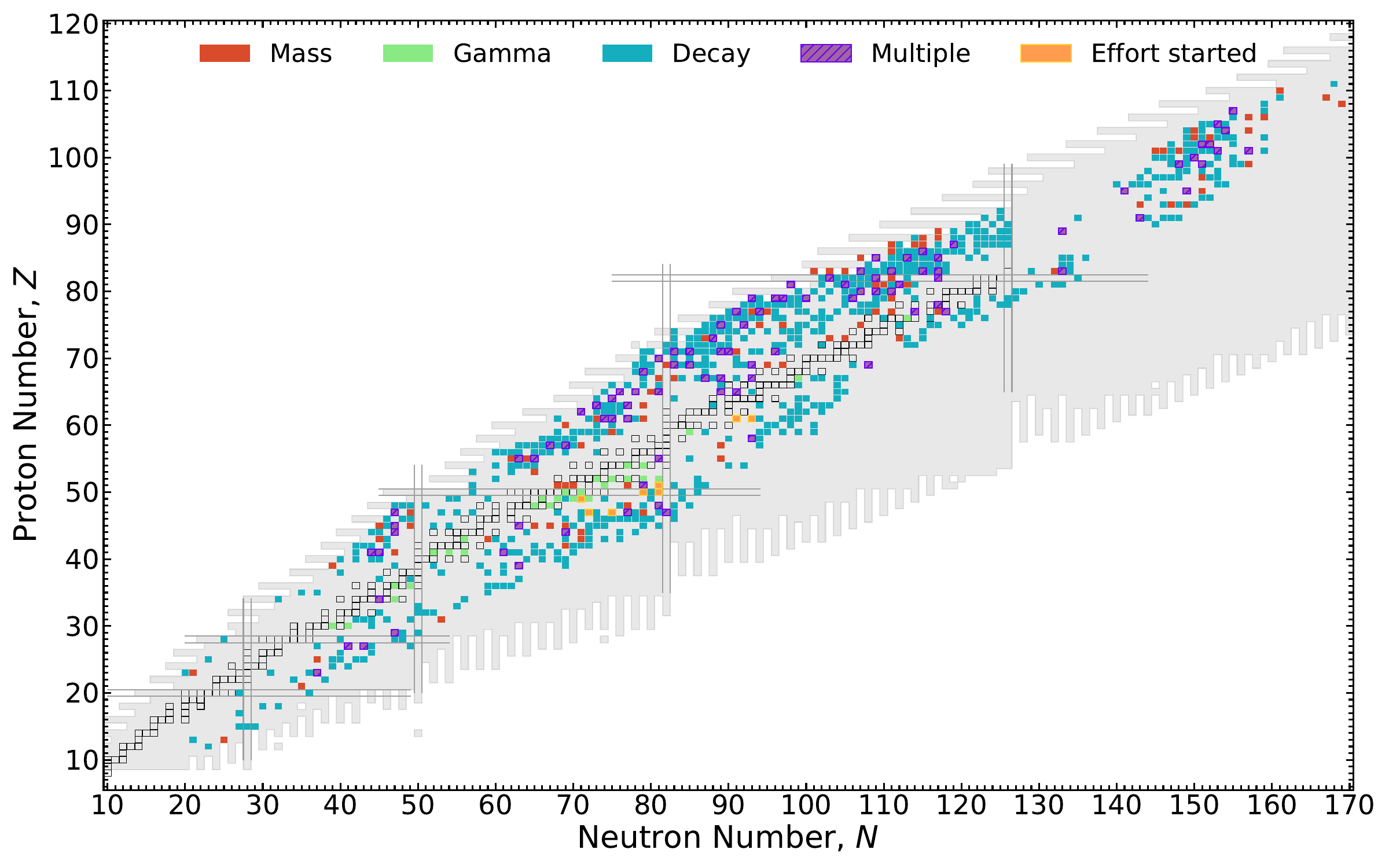}
    \caption{Unknown nuclear data contribute to uncertainties in astromer behavior. Incomplete data include level energies (Mass), transition rates (Gamma), and branching ratios (Decay). The light gray shading indicates the possible extent of bound nuclei. See text for details. }
    \label{fig:isomer_measurements}
\end{figure*}

It is important to realize that Fig. \ref{fig:isomer_measurements} was created using readily available information, and it is probably far from complete. 
Deeper investigation of known isomers and the discovery of new ones would color more squares on the chart. 


\section{Concluding remarks}

Nuclear isomers and their impact in astrophysics have long been recognized. 
Yet, the study of astromers is in its infancy. 
In addition to the inclusion of isomers in astrophysical simulations with increased fidelity, continued investment in high-precision experiments on both relatively stable and short-lived nuclei is crucial for the vitality of the field. 
Continued development of theoretical nuclear structure models that are capable of predicting isomerism in unexplored regions are requisite for guiding experimental campaigns as well as for estimating relevant quantities when data is unavailable. 
Compiling experimental and theoretical information into evaluated data suitable for rapid dissemination and application in astrophysics is imperative. 
Connecting astrophysics, nuclear structure, and reactions will lead to new insights in the future \cite{Crawford2023vision}, and astromers will play an important role.



\bmhead{Acknowledgments}

LANL is operated by Triad National Security, LLC, for the National Nuclear Security Administration of U.S. Department of Energy (Contract No. 89233218CNA000001). 
This work was partially supported by the Los Alamos National Laboratory (LANL) through its Center for Space and Earth Science (CSES). 
CSES is funded by LANL’s Laboratory Directed Research and Development (LDRD) program under project number 20240477CR-SES. 
This work was partially supported by LANL’s Laboratory Directed Research and Development (LDRD) program under project number 20230052ER. 
M.~R.~M. acknowledges support from the Directed Asymmetric Network Graphs for Research (DANGR) initiative at LANL. 


\begin{appendices}

\section{Known missing data}\label{sec:appendix}

Many isotopes with isomers are known to have incomplete data.
Table \ref{table:missing_data} summarizes missing data for nuclear \emph{isomers}, not necessarily \emph{astromers}; this information underlies Fig. \ref{fig:isomer_measurements}.
Before one can determine whether the particular state acts as an astromer, a substantial amount of data is required.
Determining whether an astromer meaningfully affects its host isotope's behavior requires even more.
This table should serve as a starting point for exploration.

\begin{center}
\begin{longtable}{c c c c c | c c c c c}
\caption[Missing data]{
  Experimental data known to be missing.
  An ``X'' denotes that the data in the corresponding category is incomplete for that isotope.
  ``Mass'' refers to the isomer energy.
  ``$\gamma$'' is important $\gamma$ transition rates identified in \citet{Misch2021sensitivity} (that work only examined neutron-rich isotopes of interest in the $r$ process), and ``Branching'' means that decay feeding from a parent nucleus is incomplete.
  The ``Started'' column indicates whether experimental work is known by the authors to have been initiated.
  }
\label{table:missing_data} \\

\hline
Isotope & Mass & $\gamma$ & Branching & Started & Isotope & Mass & $\gamma$ & Branching & Started \\
\hline 
\endfirsthead

\multicolumn{5}{c}%
{{\bfseries \tablename\ \thetable{} -- continued from previous page}} \\
\hline
Isotope & Mass & $\gamma$ & Branching & Started & Isotope & Mass & $\gamma$ & Branching & Started \\
\hline 
\endhead

\hline \multicolumn{5}{|r|}{{Continued on next page}} \\ \hline
\endfoot

\hline \hline
\endlastfoot

        $^{35}$Mg &  &  & X & No &     $^{34}$Al &  &  & X & No \\
    $^{38}$Al & X &  &  & No &     $^{42}$P &  &  & X & No \\
    $^{43}$P &  &  & X & No &     $^{44}$P &  &  & X & No \\
    $^{44}$Cl &  &  & X & No &     $^{48}$Ar &  &  & X & No \\
    $^{50}$Ar &  &  & X & No &     $^{47}$Ca &  &  & X & No \\
    $^{56}$Ca &  &  & X & No &     $^{56}$Sc & X &  &  & No \\
    $^{56}$Ti &  &  & X & No &     $^{60}$Ti &  &  & X & No \\
    $^{43}$V &  &  & X & No &     $^{44}$V & X &  &  & No \\
    $^{54}$V &  &  & X & No &     $^{59}$V &  &  & X & No \\
    $^{60}$V & X &  & X & No &     $^{63}$Cr &  &  & X & No \\
    $^{65}$Cr &  &  & X & No &     $^{48}$Mn &  &  & X & No \\
    $^{62}$Mn & X &  &  & No &     $^{64}$Mn &  &  & X & No \\
    $^{65}$Mn &  &  & X & No &     $^{67}$Mn &  &  & X & No \\
    $^{68}$Mn &  &  & X & No &     $^{66}$Fe &  &  & X & No \\
    $^{70}$Fe &  &  & X & No &     $^{64}$Co &  &  & X & No \\
    $^{68}$Co & X &  & X & No &     $^{70}$Co & X &  & X & No \\
    $^{71}$Co &  &  & X & No &     $^{76}$Co &  &  & X & No \\
    $^{53}$Ni &  &  & X & No &     $^{68}$Ni &  &  & X & No \\
    $^{75}$Ni &  &  & X & No &     $^{76}$Ni &  &  & X & No \\
    $^{76}$Cu & X &  & X & No &     $^{77}$Cu &  &  & X & No \\
    $^{79}$Cu &  &  & X & No &     $^{69}$Zn &  & X &  & No \\
    $^{71}$Zn &  & X &  & No &     $^{74}$Zn &  &  & X & No \\
    $^{73}$Ga &  &  & X & No &     $^{74}$Ga &  &  & X & No \\
    $^{75}$Ga &  &  & X & No &     $^{77}$Ga &  &  & X & No \\
    $^{80}$Ga &  &  & X & No &     $^{84}$Ga & X &  &  & No \\
    $^{77}$Ge &  &  & X & No &     $^{82}$Ge &  &  & X & No \\
    $^{83}$Ge &  &  & X & No &     $^{84}$Ge &  &  & X & No \\
    $^{83}$As &  &  & X & No &     $^{88}$As &  &  & X & No \\
    $^{66}$Se &  &  & X & No &     $^{79}$Se &  & X & X & No \\
    $^{81}$Se &  & X &  & No &     $^{90}$Se &  &  & X & No \\
    $^{70}$Br &  &  & X & No &     $^{72}$Br &  &  & X & No \\
    $^{94}$Br &  &  & X & No &     $^{83}$Kr &  & X &  & No \\
    $^{85}$Kr &  & X &  & No &     $^{95}$Kr &  &  & X & No \\
    $^{96}$Kr &  &  & X & No &     $^{97}$Kr &  &  & X & No \\
    $^{98}$Kr &  &  & X & No &     $^{82}$Rb &  &  & X & No \\
    $^{86}$Rb &  &  & X & No &     $^{100}$Rb &  &  & X & No \\
    $^{78}$Sr &  &  & X & No &     $^{91}$Sr &  &  & X & No \\
    $^{97}$Sr &  &  & X & No &     $^{99}$Sr &  &  & X & No \\
    $^{78}$Y & X &  &  & No &     $^{86}$Y &  &  & X & No \\
    $^{91}$Y &  &  & X & No &     $^{97}$Y &  &  & X & No \\
    $^{100}$Y &  &  & X & No &     $^{102}$Y & X &  & X & No \\
    $^{108}$Y &  &  & X & No &     $^{80}$Zr &  &  & X & No \\
    $^{82}$Zr &  &  & X & No &     $^{84}$Zr &  &  & X & No \\
    $^{85}$Zr &  &  & X & No &     $^{104}$Zr &  &  & X & No \\
    $^{106}$Zr &  &  & X & No &     $^{108}$Zr &  &  & X & No \\
    $^{109}$Zr &  &  & X & No &     $^{83}$Nb &  &  & X & No \\
    $^{85}$Nb & X &  & X & No &     $^{86}$Nb & X &  & X & No \\
    $^{88}$Nb & X &  &  & No &     $^{93}$Nb &  & X &  & No \\
    $^{95}$Nb &  & X &  & No &     $^{97}$Nb &  & X &  & No \\
    $^{100}$Nb &  &  & X & No &     $^{102}$Nb & X &  & X & No \\
    $^{103}$Nb &  &  & X & No &     $^{104}$Nb &  &  & X & No \\
    $^{106}$Nb &  &  & X & No &     $^{107}$Nb &  &  & X & No \\
    $^{109}$Nb &  &  & X & No &     $^{111}$Nb &  &  & X & No \\
    $^{84}$Mo &  &  & X & No &     $^{85}$Mo &  &  & X & No \\
    $^{86}$Mo &  &  & X & No &     $^{88}$Mo &  &  & X & No \\
    $^{103}$Mo &  &  & X & No &     $^{107}$Mo &  &  & X & No \\
    $^{109}$Mo &  &  & X & No &     $^{111}$Mo & X &  &  & No \\
    $^{112}$Mo &  &  & X & No &     $^{113}$Mo &  &  & X & No \\
    $^{114}$Mo &  &  & X & No &     $^{88}$Tc & X &  &  & No \\
    $^{89}$Tc &  &  & X & No &     $^{99}$Tc &  & X &  & No \\
    $^{102}$Tc & X &  &  & No &     $^{103}$Tc &  &  & X & No \\
    $^{105}$Tc &  &  & X & No &     $^{112}$Tc &  &  & X & No \\
    $^{114}$Tc & X &  &  & No &     $^{115}$Tc &  &  & X & No \\
    $^{117}$Tc &  &  & X & No &     $^{119}$Tc &  &  & X & No \\
    $^{88}$Ru &  &  & X & No &     $^{89}$Ru &  &  & X & No \\
    $^{91}$Ru & X &  & X & No &     $^{108}$Ru &  &  & X & No \\
    $^{113}$Ru & X &  & X & No &     $^{115}$Ru & X &  &  & No \\
    $^{116}$Ru &  &  & X & No &     $^{120}$Ru &  &  & X & No \\
    $^{122}$Ru &  &  & X & No &     $^{90}$Rh & X &  &  & No \\
    $^{91}$Rh &  &  & X & No &     $^{92}$Rh & X &  & X & No \\
    $^{94}$Rh & X &  &  & No &     $^{97}$Rh &  &  & X & No \\
    $^{99}$Rh &  &  & X & No &     $^{108}$Rh & X &  & X & No \\
    $^{110}$Rh & X &  &  & No &     $^{112}$Rh & X &  &  & No \\
    $^{114}$Rh & X &  &  & No &     $^{115}$Rh &  &  & X & No \\
    $^{117}$Rh &  &  & X & No &     $^{118}$Rh &  &  & X & No \\
    $^{119}$Rh &  &  & X & No &     $^{124}$Rh &  &  & X & No \\
    $^{126}$Rh &  &  & X & No &     $^{92}$Pd &  &  & X & No \\
    $^{94}$Pd &  &  & X & No &     $^{95}$Pd &  &  & X & No \\
    $^{115}$Pd &  &  & X & No &     $^{117}$Pd &  &  & X & No \\
    $^{119}$Pd &  &  & X & No &     $^{120}$Pd &  &  & X & No \\
    $^{121}$Pd &  &  & X & No &     $^{122}$Pd &  &  & X & No \\
    $^{123}$Pd &  &  & X & No &     $^{124}$Pd &  &  & X & No \\
    $^{125}$Pd &  &  & X & No &     $^{126}$Pd &  &  & X & No \\
    $^{129}$Pd &  &  & X & No &     $^{94}$Ag & X &  & X & No \\
    $^{96}$Ag & X &  &  & No &     $^{100}$Ag &  &  & X & No \\
    $^{102}$Ag &  &  & X & No &     $^{104}$Ag &  &  & X & No \\
    $^{115}$Ag &  &  & X & No &     $^{118}$Ag &  &  & X & No \\
    $^{119}$Ag & X &  & X & Yes &     $^{120}$Ag &  &  & X & No \\
    $^{122}$Ag & X &  &  & Yes &     $^{124}$Ag & X &  & X & No \\
    $^{125}$Ag &  &  & X & No &     $^{126}$Ag & X &  &  & No \\
    $^{128}$Ag &  &  & X & No &     $^{129}$Ag & X &  & X & No \\
    $^{130}$Ag &  &  & X & No &     $^{95}$Cd &  &  & X & No \\
    $^{96}$Cd &  &  & X & No &     $^{97}$Cd &  &  & X & No \\
    $^{99}$Cd &  &  & X & No &     $^{113}$Cd &  & X &  & Yes \\
    $^{115}$Cd &  & X &  & No &     $^{124}$Cd &  &  & X & No \\
    $^{125}$Cd & X &  &  & No &     $^{127}$Cd &  &  & X & No \\
    $^{129}$Cd & X &  & X & No &     $^{131}$Cd &  &  & X & No \\
    $^{133}$Cd &  &  & X & No &     $^{103}$In &  &  & X & No \\
    $^{104}$In &  &  & X & No &     $^{108}$In &  &  & X & No \\
    $^{115}$In &  & X &  & No &     $^{117}$In &  & X &  & No \\
    $^{118}$In &  &  & X & No &     $^{119}$In &  & X &  & No \\
    $^{120}$In & X &  &  & Yes &     $^{121}$In &  & X &  & No \\
    $^{123}$In &  &  & X & No &     $^{129}$In &  &  & X & No \\
    $^{133}$In &  &  & X & No &     $^{107}$Sn &  &  & X & No \\
    $^{119}$Sn &  & X &  & No &     $^{121}$Sn &  & X &  & No \\
    $^{129}$Sn &  & X &  & Yes &     $^{130}$Sn &  &  & X & No \\
    $^{131}$Sn & X &  & X & Yes &     $^{134}$Sn &  &  & X & No \\
    $^{136}$Sn &  &  & X & No &     $^{108}$Sb &  &  & X & No \\
    $^{119}$Sb & X &  &  & No &     $^{120}$Sb & X &  &  & No \\
    $^{121}$Sb & X &  &  & No &     $^{125}$Sb &  & X &  & No \\
    $^{130}$Sb &  & X & X & No &     $^{132}$Sb & X &  & X & Yes \\
    $^{134}$Sb &  &  & X & No &     $^{137}$Sb &  &  & X & No \\
    $^{138}$Sb &  &  & X & No &     $^{112}$Te &  &  & X & No \\
    $^{125}$Te &  & X &  & No &     $^{127}$Te &  & X &  & No \\
    $^{129}$Te &  & X &  & No &     $^{131}$Te &  & X &  & No \\
    $^{133}$Te &  & X &  & No &     $^{138}$Te &  &  & X & No \\
    $^{110}$I &  &  & X & No &     $^{115}$I &  &  & X & No \\
    $^{118}$I & X &  &  & No &     $^{114}$Xe &  &  & X & No \\
    $^{116}$Xe &  &  & X & No &     $^{117}$Xe &  &  & X & No \\
    $^{118}$Xe &  &  & X & No &     $^{119}$Xe &  &  & X & No \\
    $^{131}$Xe &  & X &  & No &     $^{133}$Xe &  & X &  & No \\
    $^{144}$Xe &  &  & X & No &     $^{146}$Xe &  &  & X & No \\
    $^{117}$Cs & X &  &  & No &     $^{118}$Cs & X &  & X & No \\
    $^{119}$Cs & X &  &  & No &     $^{120}$Cs & X &  & X & No \\
    $^{136}$Cs &  & X & X & No &     $^{140}$Cs &  &  & X & No \\
    $^{144}$Cs & X &  &  & No &     $^{116}$Ba &  &  & X & No \\
    $^{117}$Ba &  &  & X & No &     $^{118}$Ba &  &  & X & No \\
    $^{119}$Ba &  &  & X & No &     $^{121}$Ba &  &  & X & No \\
    $^{122}$Ba &  &  & X & No &     $^{123}$Ba &  &  & X & No \\
    $^{125}$Ba &  &  & X & No &     $^{129}$Ba &  &  & X & No \\
    $^{120}$La &  &  & X & No &     $^{121}$La &  &  & X & No \\
    $^{122}$La &  &  & X & No &     $^{123}$La &  &  & X & No \\
    $^{124}$La & X &  & X & No &     $^{125}$La &  &  & X & No \\
    $^{126}$La & X &  & X & No &     $^{127}$La &  &  & X & No \\
    $^{128}$La & X &  &  & No &     $^{146}$La & X &  &  & No \\
    $^{151}$La &  &  & X & No &     $^{124}$Ce &  &  & X & No \\
    $^{125}$Ce &  &  & X & No &     $^{126}$Ce &  &  & X & No \\
    $^{127}$Ce &  &  & X & No &     $^{128}$Ce &  &  & X & No \\
    $^{131}$Ce &  &  & X & No &     $^{148}$Ce &  &  & X & No \\
    $^{151}$Ce & X &  & X & No &     $^{152}$Ce &  &  & X & No \\
    $^{127}$Pr &  &  & X & No &     $^{129}$Pr &  &  & X & No \\
    $^{130}$Pr &  &  & X & No &     $^{131}$Pr &  &  & X & No \\
    $^{132}$Pr &  &  & X & No &     $^{133}$Pr &  &  & X & No \\
    $^{134}$Pr & X &  &  & No &     $^{135}$Pr &  &  & X & No \\
    $^{144}$Pr &  & X &  & No &     $^{146}$Pr &  &  & X & No \\
    $^{153}$Pr &  &  & X & No &     $^{154}$Pr &  &  & X & No \\
    $^{156}$Pr &  &  & X & No &     $^{158}$Pr &  &  & X & No \\
    $^{160}$Pr &  &  & X & No &     $^{129}$Nd & X &  &  & No \\
    $^{133}$Nd &  &  & X & No &     $^{135}$Nd &  &  & X & No \\
    $^{155}$Nd &  &  & X & No &     $^{156}$Nd &  &  & X & No \\
    $^{158}$Nd &  &  & X & No &     $^{161}$Nd &  &  & X & No \\
    $^{129}$Pm &  &  & X & No &     $^{133}$Pm &  &  & X & No \\
    $^{134}$Pm & X &  &  & No &     $^{135}$Pm & X &  & X & No \\
    $^{136}$Pm & X &  & X & No &     $^{138}$Pm & X &  & X & No \\
    $^{139}$Pm &  &  & X & No &     $^{140}$Pm & X &  &  & No \\
    $^{152}$Pm & X &  & X & Yes &     $^{154}$Pm & X &  &  & Yes \\
    $^{155}$Pm &  &  & X & No &     $^{158}$Pm &  &  & X & No \\
    $^{159}$Pm &  &  & X & No &     $^{160}$Pm &  &  & X & No \\
    $^{133}$Sm & X &  & X & No &     $^{135}$Sm &  &  & X & No \\
    $^{136}$Sm &  &  & X & No &     $^{137}$Sm &  &  & X & No \\
    $^{138}$Sm &  &  & X & No &     $^{160}$Sm &  &  & X & No \\
    $^{161}$Sm &  &  & X & No &     $^{162}$Sm &  &  & X & No \\
    $^{164}$Sm &  &  & X & No &     $^{136}$Eu & X &  & X & No \\
    $^{137}$Eu &  &  & X & No &     $^{138}$Eu &  &  & X & No \\
    $^{139}$Eu &  &  & X & No &     $^{140}$Eu & X &  & X & No \\
    $^{142}$Eu & X &  &  & No &     $^{151}$Eu &  &  & X & No \\
    $^{157}$Eu &  &  & X & No &     $^{160}$Eu &  &  & X & No \\
    $^{162}$Eu &  &  & X & No &     $^{164}$Eu &  &  & X & No \\
    $^{165}$Eu &  &  & X & No &     $^{166}$Eu &  &  & X & No \\
    $^{137}$Gd &  &  & X & No &     $^{139}$Gd & X &  & X & No \\
    $^{147}$Gd &  &  & X & No &     $^{163}$Gd &  &  & X & No \\
    $^{164}$Gd &  &  & X & No &     $^{168}$Gd &  &  & X & No \\
    $^{139}$Tb &  &  & X & No &     $^{141}$Tb & X &  & X & No \\
    $^{143}$Tb & X &  & X & No &     $^{145}$Tb & X &  &  & No \\
    $^{146}$Tb & X &  & X & No &     $^{154}$Tb & X &  & X & No \\
    $^{156}$Tb & X &  & X & No &     $^{158}$Tb &  &  & X & No \\
    $^{169}$Tb &  &  & X & No &     $^{170}$Tb &  &  & X & No \\
    $^{141}$Dy &  &  & X & No &     $^{143}$Dy &  &  & X & No \\
    $^{145}$Dy &  &  & X & No &     $^{147}$Dy &  &  & X & No \\
    $^{155}$Dy &  &  & X & No &     $^{165}$Dy &  &  & X & No \\
    $^{170}$Dy &  &  & X & No &     $^{146}$Ho &  &  & X & No \\
    $^{148}$Ho & X &  &  & No &     $^{150}$Ho & X &  &  & No \\
    $^{151}$Ho &  &  & X & No &     $^{153}$Ho &  &  & X & No \\
    $^{154}$Ho & X &  & X & No &     $^{155}$Ho &  &  & X & No \\
    $^{156}$Ho & X &  & X & No &     $^{158}$Ho &  &  & X & No \\
    $^{160}$Ho & X &  & X & No &     $^{166}$Ho &  & X &  & No \\
    $^{168}$Ho &  &  & X & No &     $^{169}$Ho &  &  & X & No \\
    $^{171}$Ho &  &  & X & No &     $^{172}$Ho &  &  & X & No \\
    $^{146}$Er &  &  & X & No &     $^{147}$Er & X &  & X & No \\
    $^{148}$Er &  &  & X & No &     $^{155}$Er &  &  & X & No \\
    $^{147}$Tm &  &  & X & No &     $^{148}$Tm &  &  & X & No \\
    $^{149}$Tm &  &  & X & No &     $^{151}$Tm & X &  &  & No \\
    $^{152}$Tm & X &  & X & No &     $^{153}$Tm &  &  & X & No \\
    $^{154}$Tm & X &  & X & No &     $^{155}$Tm &  &  & X & No \\
    $^{156}$Tm &  &  & X & No &     $^{157}$Tm &  &  & X & No \\
    $^{160}$Tm &  &  & X & No &     $^{161}$Tm &  &  & X & No \\
    $^{162}$Tm & X &  & X & No &     $^{164}$Tm & X &  &  & No \\
    $^{166}$Tm & X &  &  & No &     $^{175}$Tm &  &  & X & No \\
    $^{177}$Tm & X &  & X & No &     $^{149}$Yb &  &  & X & No \\
    $^{151}$Yb & X &  & X & No &     $^{153}$Yb &  &  & X & No \\
    $^{155}$Yb &  &  & X & No &     $^{157}$Yb &  &  & X & No \\
    $^{161}$Yb &  &  & X & No &     $^{150}$Lu &  &  & X & No \\
    $^{151}$Lu &  &  & X & No &     $^{153}$Lu &  &  & X & No \\
    $^{154}$Lu & X &  & X & No &     $^{155}$Lu &  &  & X & No \\
    $^{156}$Lu & X &  & X & No &     $^{157}$Lu &  &  & X & No \\
    $^{158}$Lu &  &  & X & No &     $^{159}$Lu &  &  & X & No \\
    $^{160}$Lu & X &  & X & No &     $^{161}$Lu & X &  & X & No \\
    $^{162}$Lu & X &  &  & No &     $^{164}$Lu &  &  & X & No \\
    $^{165}$Lu &  &  & X & No &     $^{167}$Lu & X &  & X & No \\
    $^{168}$Lu &  &  & X & No &     $^{154}$Hf &  &  & X & No \\
    $^{155}$Hf &  &  & X & No &     $^{157}$Hf &  &  & X & No \\
    $^{158}$Hf &  &  & X & No &     $^{159}$Hf &  &  & X & No \\
    $^{160}$Hf &  &  & X & No &     $^{161}$Hf &  &  & X & No \\
    $^{162}$Hf &  &  & X & No &     $^{168}$Hf &  &  & X & No \\
    $^{169}$Hf &  &  & X & No &     $^{171}$Hf &  &  & X & No \\
    $^{174}$Hf &  &  & X & No &     $^{185}$Hf &  &  & X & No \\
    $^{186}$Hf &  &  & X & No &     $^{156}$Ta &  &  & X & No \\
    $^{158}$Ta &  &  & X & No &     $^{159}$Ta &  &  & X & No \\
    $^{160}$Ta & X &  &  & No &     $^{161}$Ta & X &  & X & No \\
    $^{163}$Ta &  &  & X & No &     $^{165}$Ta &  &  & X & No \\
    $^{171}$Ta &  &  & X & No &     $^{172}$Ta &  &  & X & No \\
    $^{176}$Ta & X &  &  & No &     $^{178}$Ta & X &  &  & No \\
    $^{185}$Ta & X &  &  & No &     $^{186}$Ta &  &  & X & No \\
    $^{188}$Ta &  &  & X & No &     $^{157}$W &  &  & X & No \\
    $^{160}$W &  &  & X & No &     $^{161}$W &  &  & X & No \\
    $^{162}$W &  &  & X & No &     $^{163}$W &  &  & X & No \\
    $^{164}$W &  &  & X & No &     $^{165}$W &  &  & X & No \\
    $^{167}$W &  &  & X & No &     $^{169}$W &  &  & X & No \\
    $^{171}$W &  &  & X & No &     $^{172}$W &  &  & X & No \\
    $^{173}$W &  &  & X & No &     $^{175}$W &  &  & X & No \\
    $^{176}$W &  &  & X & No &     $^{189}$W &  &  & X & No \\
    $^{162}$Re &  &  & X & No &     $^{163}$Re &  &  & X & No \\
    $^{164}$Re & X &  & X & No &     $^{165}$Re &  &  & X & No \\
    $^{167}$Re & X &  & X & No &     $^{169}$Re & X &  &  & No \\
    $^{172}$Re & X &  &  & No &     $^{174}$Re &  &  & X & No \\
    $^{175}$Re &  &  & X & No &     $^{177}$Re &  &  & X & No \\
    $^{182}$Re & X &  &  & No &     $^{187}$Re &  &  & X & No \\
    $^{191}$Re &  &  & X & No &     $^{192}$Re &  &  & X & No \\
    $^{195}$Re &  &  & X & No &     $^{164}$Os &  &  & X & No \\
    $^{165}$Os &  &  & X & No &     $^{166}$Os &  &  & X & No \\
    $^{167}$Os &  &  & X & No &     $^{168}$Os &  &  & X & No \\
    $^{169}$Os &  &  & X & No &     $^{170}$Os &  &  & X & No \\
    $^{172}$Os &  &  & X & No &     $^{173}$Os &  &  & X & No \\
    $^{177}$Os &  &  & X & No &     $^{178}$Os &  &  & X & No \\
    $^{179}$Os &  &  & X & No &     $^{186}$Os &  &  & X & No \\
    $^{189}$Os &  & X &  & No &     $^{192}$Os &  &  & X & No \\
    $^{195}$Os &  &  & X & No &     $^{197}$Os &  &  & X & No \\
    $^{199}$Os &  &  & X & No &     $^{167}$Ir &  &  & X & No \\
    $^{168}$Ir & X &  & X & No &     $^{169}$Ir &  &  & X & No \\
    $^{170}$Ir & X &  &  & No &     $^{171}$Ir & X &  & X & No \\
    $^{172}$Ir & X &  &  & No &     $^{173}$Ir &  &  & X & No \\
    $^{176}$Ir &  &  & X & No &     $^{177}$Ir &  &  & X & No \\
    $^{179}$Ir &  &  & X & No &     $^{181}$Ir &  &  & X & No \\
    $^{183}$Ir &  &  & X & No &     $^{184}$Ir &  &  & X & No \\
    $^{185}$Ir &  &  & X & No &     $^{186}$Ir & X &  &  & No \\
    $^{188}$Ir & X &  &  & No &     $^{191}$Ir & X & X &  & No \\
    $^{194}$Ir & X &  &  & No &     $^{195}$Ir &  & X & X & No \\
    $^{197}$Ir &  &  & X & No &     $^{198}$Ir &  &  & X & No \\
    $^{199}$Ir &  &  & X & No &     $^{170}$Pt &  &  & X & No \\
    $^{171}$Pt &  &  & X & No &     $^{172}$Pt &  &  & X & No \\
    $^{173}$Pt &  &  & X & No &     $^{174}$Pt &  &  & X & No \\
    $^{175}$Pt &  &  & X & No &     $^{176}$Pt &  &  & X & No \\
    $^{177}$Pt &  &  & X & No &     $^{183}$Pt &  &  & X & No \\
    $^{185}$Pt &  &  & X & No &     $^{186}$Pt &  &  & X & No \\
    $^{187}$Pt &  &  & X & No &     $^{195}$Pt &  & X & X & No \\
    $^{200}$Pt &  &  & X & No &     $^{203}$Pt &  &  & X & No \\
    $^{204}$Pt &  &  & X & No &     $^{172}$Au & X &  & X & No \\
    $^{173}$Au &  &  & X & No &     $^{175}$Au & X &  & X & No \\
    $^{176}$Au & X &  & X & No &     $^{177}$Au &  &  & X & No \\
    $^{178}$Au &  &  & X & No &     $^{179}$Au & X &  & X & No \\
    $^{181}$Au &  &  & X & No &     $^{184}$Au &  &  & X & No \\
    $^{185}$Au & X &  & X & No &     $^{188}$Au &  &  & X & No \\
    $^{189}$Au &  &  & X & No &     $^{190}$Au & X &  &  & No \\
    $^{193}$Au &  &  & X & No &     $^{201}$Au &  &  & X & No \\
    $^{203}$Au &  &  & X & No &     $^{205}$Au &  &  & X & No \\
    $^{206}$Au &  &  & X & No &     $^{179}$Hg &  &  & X & No \\
    $^{181}$Hg &  &  & X & No &     $^{183}$Hg &  &  & X & No \\
    $^{185}$Hg &  &  & X & No &     $^{186}$Hg &  &  & X & No \\
    $^{187}$Hg & X &  & X & No &     $^{189}$Hg & X &  & X & No \\
    $^{190}$Hg &  &  & X & No &     $^{191}$Hg & X &  & X & No \\
    $^{193}$Hg &  &  & X & No &     $^{207}$Hg &  &  & X & No \\
    $^{208}$Hg &  &  & X & No &     $^{179}$Tl & X &  & X & No \\
    $^{183}$Tl &  &  & X & No &     $^{185}$Tl &  &  & X & No \\
    $^{186}$Tl & X &  & X & No &     $^{187}$Tl &  &  & X & No \\
    $^{188}$Tl &  &  & X & No &     $^{189}$Tl &  &  & X & No \\
    $^{190}$Tl & X &  &  & No &     $^{191}$Tl & X &  &  & No \\
    $^{192}$Tl &  &  & X & No &     $^{193}$Tl & X &  & X & No \\
    $^{194}$Tl & X &  &  & No &     $^{195}$Tl &  &  & X & No \\
    $^{211}$Tl &  &  & X & No &     $^{213}$Tl &  &  & X & No \\
    $^{214}$Tl &  &  & X & No &     $^{184}$Pb &  &  & X & No \\
    $^{185}$Pb & X &  & X & No &     $^{186}$Pb &  &  & X & No \\
    $^{187}$Pb &  &  & X & No &     $^{188}$Pb &  &  & X & No \\
    $^{189}$Pb &  &  & X & No &     $^{190}$Pb &  &  & X & No \\
    $^{191}$Pb & X &  & X & No &     $^{193}$Pb & X &  &  & No \\
    $^{195}$Pb &  &  & X & No &     $^{199}$Pb & X &  & X & No \\
    $^{217}$Pb &  &  & X & No &     $^{184}$Bi & X &  &  & No \\
    $^{186}$Bi & X &  &  & No &     $^{188}$Bi & X &  &  & No \\
    $^{190}$Bi & X &  & X & No &     $^{191}$Bi &  &  & X & No \\
    $^{192}$Bi &  &  & X & No &     $^{193}$Bi &  &  & X & No \\
    $^{194}$Bi & X &  & X & No &     $^{195}$Bi &  &  & X & No \\
    $^{196}$Bi &  &  & X & No &     $^{197}$Bi &  &  & X & No \\
    $^{198}$Bi & X &  & X & No &     $^{199}$Bi &  &  & X & No \\
    $^{200}$Bi & X &  & X & No &     $^{201}$Bi &  &  & X & No \\
    $^{212}$Bi &  &  & X & No &     $^{215}$Bi & X &  &  & No \\
    $^{216}$Bi & X &  & X & No &     $^{217}$Bi &  &  & X & No \\
    $^{192}$Po &  &  & X & No &     $^{194}$Po &  &  & X & No \\
    $^{195}$Po &  &  & X & No &     $^{196}$Po &  &  & X & No \\
    $^{197}$Po &  &  & X & No &     $^{198}$Po &  &  & X & No \\
    $^{199}$Po &  &  & X & No &     $^{200}$Po &  &  & X & No \\
    $^{201}$Po &  &  & X & No &     $^{202}$Po &  &  & X & No \\
    $^{217}$Po &  &  & X & No &     $^{218}$Po &  &  & X & No \\
    $^{192}$At & X &  &  & No &     $^{194}$At & X &  & X & No \\
    $^{196}$At &  &  & X & No &     $^{197}$At &  &  & X & No \\
    $^{198}$At & X &  & X & No &     $^{199}$At &  &  & X & No \\
    $^{200}$At &  &  & X & No &     $^{201}$At &  &  & X & No \\
    $^{202}$At & X &  & X & No &     $^{206}$At &  &  & X & No \\
    $^{208}$At &  &  & X & No &     $^{218}$At &  &  & X & No \\
    $^{219}$At &  &  & X & No &     $^{221}$At &  &  & X & No \\
    $^{197}$Rn & X &  &  & No &     $^{198}$Rn &  &  & X & No \\
    $^{199}$Rn &  &  & X & No &     $^{200}$Rn &  &  & X & No \\
    $^{201}$Rn & X &  & X & No &     $^{202}$Rn &  &  & X & No \\
    $^{203}$Rn &  &  & X & No &     $^{204}$Rn &  &  & X & No \\
    $^{205}$Rn &  &  & X & No &     $^{206}$Rn &  &  & X & No \\
    $^{208}$Rn &  &  & X & No &     $^{198}$Fr & X &  &  & No \\
    $^{199}$Fr &  &  & X & No &     $^{201}$Fr & X &  &  & No \\
    $^{202}$Fr & X &  &  & No &     $^{204}$Fr &  &  & X & No \\
    $^{206}$Fr & X &  & X & No &     $^{207}$Fr &  &  & X & No \\
    $^{209}$Fr &  &  & X & No &     $^{210}$Fr &  &  & X & No \\
    $^{211}$Fr &  &  & X & No &     $^{212}$Fr &  &  & X & No \\
    $^{213}$Fr &  &  & X & No &     $^{202}$Ra &  &  & X & No \\
    $^{203}$Ra & X &  &  & No &     $^{205}$Ra & X &  &  & No \\
    $^{207}$Ra &  &  & X & No &     $^{208}$Ra &  &  & X & No \\
    $^{210}$Ra &  &  & X & No &     $^{211}$Ra &  &  & X & No \\
    $^{212}$Ra &  &  & X & No &     $^{213}$Ra &  &  & X & No \\
    $^{214}$Ra &  &  & X & No &     $^{206}$Ac & X &  &  & No \\
    $^{208}$Ac &  &  & X & No &     $^{210}$Ac &  &  & X & No \\
    $^{211}$Ac &  &  & X & No &     $^{212}$Ac &  &  & X & No \\
    $^{214}$Ac &  &  & X & No &     $^{215}$Ac &  &  & X & No \\
    $^{222}$Ac & X &  & X & No &     $^{211}$Th &  &  & X & No \\
    $^{212}$Th &  &  & X & No &     $^{214}$Th &  &  & X & No \\
    $^{216}$Th &  &  & X & No &     $^{235}$Th &  &  & X & No \\
    $^{214}$Pa &  &  & X & No &     $^{216}$Pa &  &  & X & No \\
    $^{226}$Pa &  &  & X & No &     $^{234}$Pa & X &  & X & No \\
    $^{235}$Pa &  &  & X & No &     $^{237}$Pa &  &  & X & No \\
    $^{238}$Pa &  &  & X & No &     $^{239}$Pa &  &  & X & No \\
    $^{217}$U &  &  & X & No &     $^{236}$Np & X &  &  & No \\
    $^{239}$Np &  &  & X & No &     $^{240}$Np & X &  &  & No \\
    $^{241}$Np &  &  & X & No &     $^{242}$Np & X &  &  & No \\
    $^{243}$Np &  &  & X & No &     $^{238}$Pu &  &  & X & No \\
    $^{245}$Pu &  &  & X & No &     $^{246}$Pu &  &  & X & No \\
    $^{236}$Am & X &  & X & No &     $^{241}$Am &  &  & X & No \\
    $^{244}$Am & X &  & X & No &     $^{246}$Am & X &  &  & No \\
    $^{236}$Cm &  &  & X & No &     $^{238}$Cm &  &  & X & No \\
    $^{239}$Cm &  &  & X & No &     $^{240}$Cm &  &  & X & No \\
    $^{249}$Cm &  &  & X & No &     $^{250}$Cm &  &  & X & No \\
    $^{240}$Bk &  &  & X & No &     $^{242}$Bk &  &  & X & No \\
    $^{243}$Bk &  &  & X & No &     $^{244}$Bk &  &  & X & No \\
    $^{245}$Bk &  &  & X & No &     $^{248}$Bk & X &  &  & No \\
    $^{249}$Bk &  &  & X & No &     $^{250}$Bk &  &  & X & No \\
    $^{242}$Cf &  &  & X & No &     $^{246}$Cf &  &  & X & No \\
    $^{247}$Cf &  &  & X & No &     $^{249}$Cf &  &  & X & No \\
    $^{251}$Cf &  &  & X & No &     $^{246}$Es &  &  & X & No \\
    $^{247}$Es & X &  & X & No &     $^{248}$Es &  &  & X & No \\
    $^{250}$Es & X &  & X & No &     $^{251}$Es &  &  & X & No \\
    $^{254}$Es &  &  & X & No &     $^{255}$Es &  &  & X & No \\
    $^{256}$Es & X &  &  & No &     $^{245}$Fm &  &  & X & No \\
    $^{246}$Fm &  &  & X & No &     $^{247}$Fm &  &  & X & No \\
    $^{249}$Fm &  &  & X & No &     $^{250}$Fm & X &  & X & No \\
    $^{253}$Fm &  &  & X & No &     $^{255}$Fm &  &  & X & No \\
    $^{246}$Md & X &  &  & No &     $^{247}$Md & X &  &  & No \\
    $^{248}$Md &  &  & X & No &     $^{249}$Md & X &  &  & No \\
    $^{250}$Md &  &  & X & No &     $^{251}$Md &  &  & X & No \\
    $^{252}$Md &  &  & X & No &     $^{253}$Md &  &  & X & No \\
    $^{254}$Md & X &  & X & No &     $^{258}$Md & X &  & X & No \\
    $^{260}$Md &  &  & X & No &     $^{249}$No &  &  & X & No \\
    $^{251}$No &  &  & X & No &     $^{252}$No &  &  & X & No \\
    $^{253}$No & X &  & X & No &     $^{254}$No & X &  & X & No \\
    $^{257}$No &  &  & X & No &     $^{252}$Lr &  &  & X & No \\
    $^{253}$Lr & X &  &  & No &     $^{254}$Lr &  &  & X & No \\
    $^{255}$Lr & X &  &  & No &     $^{256}$Lr &  &  & X & No \\
    $^{257}$Lr &  &  & X & No &     $^{258}$Lr &  &  & X & No \\
    $^{262}$Lr &  &  & X & No &     $^{253}$Rf &  &  & X & No \\
    $^{254}$Rf & X &  &  & No &     $^{255}$Rf &  &  & X & No \\
    $^{257}$Rf &  &  & X & No &     $^{258}$Rf & X &  & X & No \\
    $^{261}$Rf & X &  &  & No &     $^{256}$Db &  &  & X & No \\
    $^{257}$Db &  &  & X & No &     $^{258}$Db & X &  & X & No \\
    $^{259}$Db &  &  & X & No &     $^{261}$Sg &  &  & X & No \\
    $^{263}$Sg & X &  &  & No &     $^{265}$Sg & X &  &  & No \\
    $^{262}$Bh & X &  & X & No &     $^{266}$Bh &  &  & X & No \\
    $^{267}$Hs &  &  & X & No &     $^{277}$Hs & X &  &  & No \\
    $^{270}$Mt &  &  & X & No &     $^{276}$Mt & X &  &  & No \\
    $^{271}$Ds & X &  &  & No &     $^{279}$Rg &  &  & X & No \\
    $^{283}$Cn &  &  & X & No &  \\

\end{longtable}
\end{center}

\end{appendices}


\bibliography{astromers_review}

\end{document}